\documentclass[aps,pre,showpacs,twocolumn]{revtex4}
\usepackage{bm}
\usepackage{graphicx}
\bibstyle{approve.bib}
\usepackage{amssymb}
\usepackage{amsmath}
\usepackage{esint}
\usepackage{epstopdf}
\usepackage{color}
\usepackage{gensymb}
\usepackage{mathtools}
\usepackage{suffix}
\usepackage{fancyhdr}

\newcommand{\be}{\begin{equation}}
\newcommand{\ee}{\end{equation}}
\newcommand{\bea}{\begin{eqnarray}}
\newcommand{\eea}{\end{eqnarray}}
\newcommand{\ph}{\mathbf{\Psi}}
\newcommand{\lan}{\left\langle}
\newcommand{\ran}{\right\rangle}
\newcommand{\br}{\mathbf{r}}
\newcommand{\bR}{\mathbf{R}}

\newcommand{\hk}{\hat{k}}

\newcommand{\bq}{\mathbf{q}}

\newcommand{\e}{\varepsilon}

\newcommand{\tv}{\tilde{v}}
\newcommand{\tq}{\tilde{Q}}

\newcommand{\may}{\tilde{h}_{ij}(q)}

\newcommand{\tchi}{\tilde{\chi}}
\newcommand{\tG}{\tilde{G}}
\newcommand{\tgg}{\tilde{{\mathcal G}}}

\newcommand{\tep}{\tilde{\varepsilon}}

\newcommand{\m}{_{\rm m}}
\newcommand{\ce}{_{\rm c}}
\newcommand{\G}{_{\rm G}}

\newcommand{\h}{_{\rm h}}
\newcommand{\lo}{_{\rm l}}
\newcommand{\n}{_{\rm n}}

\newcommand{\s}{_{\rm s}}

\newcommand{\hn}{\hat{n}}

\newcommand{\ew}{\varepsilon_{\rm w}}
\newcommand{\SB}[1]{\textcolor{black} {#1}}

\begin{document}

\title{Self-consistent electrostatic formalism of bulk electrolytes based on the asymmetric treatment of the short- and long-range ion interactions}

\author{Sahin Buyukdagli}
\address{Department of Physics, Bilkent University, Ankara 06800, Turkey}

\begin{abstract}

\noindent{\bf ABSTRACT:}  We predict the thermodynamic behavior of bulk electrolytes from an ionic Hard-Core (HC) size-augmented self-consistent formalism incorporating asymmetrically the short- and long-range ion interactions via their virial and cumulant treatment, respectively. The characteristic splitting length separating these two ranges is obtained from a variational equation solved together with the Schwinger-Dyson (SD) equations. Via comparison with simulation results from the literature, we show that the asymmetric treatment of the distinct interaction ranges significantly extends the validity regime of our previously developed purely cumulant-level Debye-H\"{u}ckel (DH) theory. Namely, for monovalent solutions with typical ion sizes, the present formalism can accurately predict \SB{up to molar concentrations the liquid pressure dominated by HC interactions, the  internal energies driven by charge correlations, and the local ion distributions governed by the competition between HC and electrostatic interactions}. We evaluate as well the screening length of the liquid and investigate the deviations of the \SB{macromolecular} interaction range from the DH length. \SB{In fair agreement with simulations and experiments, our theory is shown to reproduce the overscreening and underscreening effects occurring respectively  in submolar mono- and multivalent electrolytes}.
\end{abstract}

\date{\today}
\maketitle   

\section{Introduction}

The local violations of the electroneutrality condition always maintained above the characteristic scale of a few nanometers play a direct role in the continuity of life on Earth. From ion transport through plasma membranes~\cite{Bont,biomatter} to viral infection~\cite{gn1} and DNA-histone interactions essential to chromatin stability~\cite{gn1}, attractive electrostatic forces balanced by HC repulsion regulate numerous nanoscale processes vital to living organisms. In addition, the collective action of the electrostatic and HC interactions coupled with fluid dynamics lies at the heart of various artificially induced processes such as electrokinetic energy conversion~\cite{Boc1,Boc2}, water purification and desalination~\cite{Yar,Szymczyk}, and nanopore-based DNA sequencing~\cite{Wanunu}. Thus, the comprehension and control of these phenomena require the accurate formulation of the many-body charge interactions at the nanoscale. 

The DH formalism introduced a century ago has been the leading theory enabling the consistent characterization of the thermodynamics of bulk electrolytes~\cite{DH}. Despite its success as a pioneering formulation of ion interactions, the original DH formalism suffers from two major limitations. First, the theory incorporating the hydrated ion size as a screening-free cavity neglects the repulsive pairwise HC interactions. Then, the possibility to recover the formalism from the one-loop-level expansion of the liquid partition function indicates its electrostatic weak-coupling (WC) nature~\cite{Buyuk2024}. As a result, the validity of the theory is limited to the characterization of weakly coupled monovalent salts at dilute concentrations.

The access to the opposite regime of multivalent ions and condensed solutions has been provided by Monte-Carlo (MC) simulations enabling the exact characterization of the liquid thermodynamics~\cite{Valleau,Svensson,NetzMC,NetzMC2,LevinMC}. Dense liquids of high ion valency have been also investigated by approximate thermodynamic formalisms such as density functional and integral equation theories~\cite{Blum,Henderson,Boda,Hoye,Hansen,Roji}.

In order to gain analytical insight into the thermodynamics of charged liquids, the aforementioned approaches should be complemented by mathematically transparent theories of reduced numerical complexity. Along these lines, Attard~\cite{AttardPRE,AttardJCP,AttardRev} and Kjellander~\cite{Kj1,Kj2,Kj1995,KjJCP2016,Kj2020} developed effective electrostatic theories based on the forcing of the zeroth and second moment conditions on DH-like charge distribution functions. On the side of the conventional statistical mechanics, the functional integral formulation of the liquid partition function~\cite{Kholodenko1,Kholodenko2,PodWKB} has been successfully used for the characterization of the liquid thermodynamics via systematic perturbation techniques~\cite{NetzLoop} and variational methods~\cite{netzvar,HatloVar0,Buyuk2012} in confined media.

The predictions of these WC-level functional integral theories accurate for monovalent ions have been also extended to the strong-coupling regime of multivalent charges by virial expansion techniques~\cite{NetzSC} and hybrid formalisms treating the electrolyte components asymmetrically according to their individual valency~\cite{Podgornik2011,Buyuk2020}. Moreover, Chen et al.~\cite{sp,Chen} and  Santangelo~\cite{Santangelo} developed efficient splitting techniques based on the asymmetric treatment of the ionic short- and long-wavelength interactions whose ranges are separated by a characteristic splitting length adjusted arbitrarily. Then, Santangelo's splitting theory has been ingeniously upgraded by Hatlo and Lue's variational formalism enabling the explicit calculation of the splitting parameter via the numerical minimization of the electrostatic grand potential~\cite{HatloSoft,HatloEpl,LueSoft}. The corresponding theory treating the long-range interactions at the mean-field-level and the short-range interactions within a cumulant expansion scheme has been shown to reproduce self-consistently the local density and pressure of confined electrolytes from weak to strong electrostatic coupling regime.

As the competition between the electrostatic and HC interactions emerges at submolar ion concentrations~\cite{Book}, the accuracy of the functional integral models neglecting the interionic HC coupling is limited to dilute salt solutions. With the aim to overcome this limitation, we have recently developed a calculation scheme enabling the calculation of the liquid thermodynamic functions by explicitly incorporating the ionic HC size~\cite{Buyuk2024}. It should be noted that the underlying cumulant expansion around a gaussian reference Hamiltonian corresponds to an electrostatic WC approximation. Consequently, while the resulting cumulant-corrected DH (CCDH) theory can predict the HC-dominated liquid pressure up to molar concentrations, its predictions for the electrostatically driven internal energy are  limited to submolar concentrations. 

In order to extend the electrostatic coupling regime covered by the CCDH formalism, in this article, we develop a self-consistent DH (SCDH) theory of bulk electrolytes based on the asymmetric treatment of the short- and long-range electrostatic fluctuations. In Sec.~\ref{mod}, we derive the field-theoretic partition function of the liquid embodying the corresponding splitting together with the pairwise HC interactions, and obtain a new variational equation solved by the splitting parameter separating these distinct interaction ranges. Then, we derive the SD equations relating this variational identity to the physical parameters of the system. Finally, we calculate the total correlation function required for the computation of the thermodynamic functions and the screening length of the electrolyte. The thermodynamic averages involved in these formally exact identities are evaluated via an hybrid calculation scheme that consists in incorporating asymmetrically the electrostatic fluctuations of short- and long-wavelengths  within the virial and cumulant approximations, respectively. This scheme corresponding to a variational upgrade of our electrostatic WC-level CCDH formalism~\cite{Buyuk2024} allows to avoid the cumulant-level WC treatment of the short-range ion interactions. 

In Sec.~\ref{res}, we compare our predictions with hypernetted-chain (HNC) and MC data for various \SB{ion sizes, valencies, and concentrations}. We find that the asymmetric variational treatment of the distinct interaction modes significantly improves the accuracy of the CCDH formalism in predicting \SB{the HC-dominated osmotic pressures, the electrostatic correlation-driven internal energies, and the ion pair distributions governed by the competing HC and electrostatic interactions}. This upgrade is the main progress of our work. Via additional comparison with simulations and experiments, we equally show that our approach can also reproduce with reasonable accuracy the underscreening and overscreening effects occurring in mono- and multivalent solutions.

\section{Model and theory}

\label{mod}

\subsection{Liquid partition function}

We introduce here the electrolyte model and derive the partition function of the solution. The solvent-implicit bulk electrolyte is composed of $p$ ion species. The ion charge of the species $i$ with valency $q_i$ and concentration $n_i$ is placed at the center of an impenetrable HC sphere of radius $d$ corresponding to its hydrated volume. The electrolyte of temperature $T$ has dielectric permittivity $\ew\e_0$ , where $\e_0$ and $\ew$ are the vacuum and relative water permittivities, respectively.

The grand canonical partition function of the liquid involving the fugacity $\lambda_i$ of the ion species $i$ reads
\bea
\label{eq1}
Z\G=\prod_{i=1}^p\sum_{N_i=1}^\infty\frac{\lambda_i^{N_i}}{N_i!}\prod_{j=1}^p\prod_{k=1}^{N_j}\int\mathrm{d}^3\br_{jk}e^{-\beta(E\m+E\n)},
\eea
where the pairwise interaction energy 
\bea
\label{eq2}
\beta E\m&=&\frac{1}{2}\int\mathrm{d^3}\br\mathrm{d^3}\br'\left[\hn\ce(\br)v\ce(\br-\br')\hn\ce(\br')\right.\\
&&\left.\hspace{2cm}+\hn\h(\br)v\h(\br-\br')\hn\h(\br')\right]\nonumber
\eea
has been expressed in terms of the ionic charge and number density operators defined as
\bea
\label{eq3I}
\hn\ce(\br)&=&\sum_{i=1}^p\sum_{j=1}^{N_i}q_i\delta^3(\br-\br_{ij})+\sum_{k=1}^{N_{\rm m}}C_k\delta^3(\br-\bR_k);\\
\label{eq3II}
\hn\h(\br)&=&\sum_{i=1}^p\sum_{j=1}^{N_i}\delta^3(\br-\br_{ij}).
\eea
In Eq.~(\ref{eq2}), the electrostatic charge interactions are mediated by the inverse of the bulk Coulomb operator
\be\label{eq4}
v\ce^{-1}(\br,\br')=-\frac{k_{\rm B}T\ew\e_0}{e^2}\nabla^2\delta^3(\br-\br')
\ee
corresponding to the standard Coulomb potential 
\be
\label{co}
v\ce(\br,\br')=\frac{\ell_{\rm B}}{||\br-\br'||}, 
\ee
where $\ell_{\rm B}=e^2/(4\pi\ew\e_0 k_{\rm B}T)$ is the Bjerrum length defined in terms of the electron charge $e$ and the Boltzmann constant $k_{\rm B}$. Eq.~(\ref{eq2}) also includes the pairwise HC interaction potential $v\h(r)$ defined as
\be\label{eq5}
e^{-v\h(\br)}=\theta(r-d),
\ee
where $\theta(x)$ is the Heaviside step function~\cite{math}. Finally, in Eq.~(\ref{eq3I}), we introduced $N_{\rm m}$ fixed charges of valencies $C_k$ that will allow to derive the net correlation function.

Eq.~(\ref{eq1}) contains as well the single-body energy 
\be
\label{eq6}
\beta E\n=\sum_{i=1}^p\int\mathrm{d}^3\br\;w_i(\br)\sum_{j=1}^{N_i}\delta^3(\br-\br_{ij})-\sum_{i=1}^pN_i\epsilon_i
\ee
including the potential $w_i(\br_{ij})$ introduced for the derivation of the average ion densities, and the ionic self-energy
\be\label{eq7}
\epsilon_i=\frac{1}{2}\left[q_i^2v\ce(0)+v\h(0)\right]
\ee
that has been subtracted from the Hamiltonian. 

Following the approach of Refs.~\cite{Santangelo,HatloEpl,HatloSoft}, we split now the Coulomb potential into a short-range and a long-range component,
\be\label{eq8}
v\ce(\br,\br')=v\s(\br,\br')+v\lo(\br,\br'),
\ee
whose functional forms will be specified  below. In the first term on the r.h.s. of Eq.~(\ref{eq2}), this splitting gives rise to two types of electrostatic energy components. Thus, introducing in Eq.~(\ref{eq1}) an Hubbard-Stratonovich transformation for each type of pairwise interaction,
\bea\label{eq9}
\hspace{-5mm}&&e^{-\frac{1}{2}\int\mathrm{d}^3\br\mathrm{d}^3\br'\hn_\gamma(\br)v_\gamma(\br-\br')\hn_\gamma(\br')}\\
\hspace{-5mm}&&=\int\mathcal{D}\psi_\gamma\;e^{-\frac{1}{2}\int\mathrm{d}^3\br\mathrm{d}^3\br'\psi_\gamma(\br)v^{-1}_\gamma(\br-\br')\psi_\gamma(\br')}e^{i\int\mathrm{d}^3\br \hn_\gamma(\br)\psi_\gamma(\br)},\nonumber
\eea
where $\psi_\gamma(\br)$ stand for the auxiliary potentials associated with the short-range ($\gamma={\rm s}$) and long-range ($\gamma={\rm l}$) Coulomb interactions, and the pairwise HC interactions ($\gamma={\rm h}$), one can recast the grand-canonical partition function as a functional integral over these potentials,
\be
\label{eq10}
Z_{\rm G}=\int\frac{\mathcal{D}\ph}{\sqrt{{\rm det}\left[v\s v\lo v\h\right]}}\;e^{-\beta H[\ph]}.
\ee
In Eq.~(\ref{eq10}), we introduced the shorthand vector notations for the fluctuating potentials $\ph=(\psi\s,\psi\lo,\psi\h)$ and the functional integration measure $\mathcal{D}\ph=\mathcal{D}\psi\s\mathcal{D}\psi\lo\mathcal{D}\psi\h$, and defined the Hamiltonian functional 
\bea\label{eq11}
\beta H[\ph]&=&\sum_{\gamma={\rm s,l,h}}\int\frac{\mathrm{d}^3\br\mathrm{d}^3\br'}{2}\psi_\gamma(\br)v_\gamma^{-1}(\br,\br')\psi_\gamma(\br')\\
&&-\sum_{i=1}^p\lambda_i \int\mathrm{d}^3\br\;\hk_i(\br)-i\sum_{k=1}^{N_{\rm m}}C_k\left[\psi\lo+\psi\s\right]_{\bR_k}\nonumber
\eea
including the fluctuating ion density function
\be
\label{eq12}
\hk_i(\br)=e^{\epsilon_i-w_i(\br)+i\psi\h(\br)+iq_i\left[\psi\s(\br)+\psi\lo(\br)\right]}.
\ee
Unless stated otherwise, in the remainder, we will omit the macromolecular charges and set $C_k=0$.

\subsection{Splitting scheme}

\subsubsection{Short- and long-range interaction potentials}

In this article, the internal energy and pressure of the electrolyte will be calculated with the long-range component of the potential~(\ref{eq8}) chosen as the inverse of the following operator originally introduced in Ref.~\cite{HatloSoft},
\be
\label{vl1}
v^{-1}\lo(\br,\br')=\left(1-\sigma^2\nabla^2+\sigma^4\nabla^4\right)v\ce^{-1}(\br,\br'),
\ee
where $\sigma$ stands for the auxiliary splitting length separating the short- and long-range electrostatic interactions. In our article, the Fourier transform (FT)  and the inverse FT of the general function $f(\br)$ are respectively defined as $\tilde{f}(q)=\int\mathrm{d}^3\br f(\br)e^{-i\bq\cdot\br}$ and $f(\br)=(2\pi)^{-3}\int\mathrm{d}^3\br \tilde{f}(q)e^{i\bq\cdot\br}$. Inverting now Eq.~(\ref{vl1}) in Fourier space, and using the FT of the Coulomb potential $\tv\ce(q)=4\pi\ell_{\rm B}/q^2$ together with the constraint~(\ref{eq8}), the FT of the short- and long-wavelength potentials follow as
\bea
\label{vl2}
\tv_{\rm s}(q)&=&\frac{4\pi\ell_{\rm B}}{q^2}\frac{\sigma^2q^2+\sigma^4q^4}{1+\sigma^2q^2+\sigma^4q^4};\\
\label{vl3}
\tv_{\rm l}(q)&=&\frac{4\pi\ell_{\rm B}}{q^2}\left(1+\sigma^2q^2+\sigma^4q^4\right)^{-1}.
\eea
Finally, the inverse FT of Eqs.~(\ref{vl2})-(\ref{vl3}) yield 
\bea
\label{vl4}
v_{\rm s}(r)&=&\frac{\ell_{\rm B}}{r}\left\{\cos\left(\frac{r}{2\sigma}\right)+\frac{1}{\sqrt{3}}\sin\left(\frac{r}{2\sigma}\right)\right\}e^{-\frac{\sqrt{3}r}{2\sigma}};\\
\label{vl4II}
v_{\rm l}(r)&=&\frac{\ell_{\rm B}}{r}-v_{\rm s}(r).
\eea

\subsubsection{Variational equation for the splitting parameter}

The value of the splitting length $\sigma$ required for the explicit calculation of the thermodynamic functions can be determined by exploiting the invariance of the partition function~(\ref{eq1}) and the grand-potential $\Omega\G=-k_{\rm B}T\ln Z\G$ under the alteration of this parameter, i.e. $\partial_{\sigma}\Omega\G=0$. Expressing the latter equality via the functional integral representation~(\ref{eq10}) of the partition function, the variational equation satisfied by the parameter $\sigma$ follows as
\be\label{eq13}
\sum_{\gamma=\{{\rm s,l}\}}\int\mathrm{d}^3\br\mathrm{d}^3\br'\left[\partial_{\sigma}v_\gamma^{-1}(\br,\br')\right]\left\{G_\gamma(\br,\br')-v_\gamma(\br,\br')\right\}=0.
\ee
In Eq.~(\ref{eq13}), the two-point correlation functions
\be
\label{eq14}
G_\gamma(\br,\br')=\lan\psi_\gamma(\br)\psi_\gamma(\br')\ran
\ee
involve the functional averages over the fluctuating potentials $\psi_\gamma(\br)$ defined for a general functional $F[\ph]$ as
\be\label{eq15}
\lan F[\ph]\ran=\frac{1}{Z_{\rm G}}\int\mathcal{D}\ph\;e^{-\beta H[\ph]}F[\ph].
\ee

The formally exact variational Eq.~(\ref{eq13}) that will enable the asymmetric treatment of the distinct interaction modes is one of the key results of the present work.  This identity equally valid for confined liquids can be also extended directly to more general splitting potentials involving multiple splitting parameters $\sigma_i$. In the present case of bulk liquids where the translational symmetry implies $v_\gamma(\br,\br')=v_\gamma(\br-\br')$ and $G_\gamma(\br,\br')=G_\gamma(\br-\br')$, Eq.~(\ref{eq13}) can be expressed in Fourier space as
\bea
\label{eq16}
\sum_{\gamma=\{{\rm s,l}\}}\int_0^\infty\mathrm{d}qq^2\left\{\tG_\gamma(q)-\tv_\gamma(q)\right\}\partial_\sigma\tv_\gamma(q)=0.
\eea

\subsection{Electrostatic SD identities}

The correlation functions~(\ref{eq14}) can be related to the physical parameters of the charged liquid via the electrostatic SD identities~\cite{Buyuk2024}. In order to extend the derivation of these identities to multiple electrostatic potential components, we define first the functional integral
\be\label{eq17}
J=\int\mathcal{D}\ph\;e^{-\beta H[\ph]}F[\ph].
\ee 
Then, we introduce an infinitesimal shift of the electrostatic potential component $\psi_\gamma(\br)$ for $\gamma=\{s,l\}$ as $\psi_\gamma(\br)\to\psi_\gamma(\br)+\delta\psi_\gamma(\br)$, and expand  Eq.~(\ref{eq17}) at the linear order in the function $\delta\psi_\gamma(\br)$. The resulting variation of the integral~(\ref{eq17}) follows as
\bea\label{eq18}
\delta J&=&\int\mathrm{d}\br\delta\psi_\gamma(\br)\int\mathcal{D}\ph e^{-\beta H[\ph]}\\
&&\hspace{2.3cm}\times\left\{\frac{\delta F[\ph]}{\delta\psi_\gamma(\br)}-F[\ph]\frac{\delta H[\ph]}{\delta\psi_\gamma(\br)}\right\}.\nonumber
\eea
At this point, we take into account the invariance of the functional integral~(\ref{eq17}) under this potential shift, and set Eq.~(\ref{eq18}) to zero~\cite{justin}. Dividing the resulting expression by the partition function~(\ref{eq10}), one obtains the identity
\be
\label{eq19}
\lan\frac{\delta F[\ph]}{\delta\psi_\gamma(\br)}\ran=\lan F[\ph]\frac{\delta H[\ph]}{\delta\psi_\gamma(\br)}\ran.
\ee
Setting in Eq.~(\ref{eq19}) $F[\ph]=1$ and $F[\ph]=\psi_\gamma(\br')$, the SD identities respectively follow in the form
\be\label{eq20}
\lan\frac{\delta\left(\beta H[\ph]\right)}{\delta\psi_\gamma(\br)}\ran=0;\hspace{3mm}\lan\frac{\delta\left(\beta H[\ph]\right)}{\delta\psi_\gamma(\br)}\psi_\gamma(\br')\ran=\delta^3(\br-\br').
\ee

The substitution of the Hamiltonian~(\ref{eq11}) into the second SD identity in Eq.~(\ref{eq20}) now yields
\bea
\label{eq21}
&&\hspace{-3mm}\int\mathrm{d}^3\br_1 v_\gamma^{-1}(\br,\br_1)G_\gamma(\br_1,\br')-i\sum_{i=1}^p\lambda_iq_i \lan\hk_i(\br)\psi_\gamma(\br')\ran\nonumber\\
&&\hspace{-3mm}=\delta^3(\br-\br').
\eea
Finally, using the equality 
\be\label{inv}
\int\mathrm{d}^3\br_1v_\gamma^{-1}(\br,\br_1)v_\gamma(\br_1,\br')=\delta^3(\br-\br'), 
\ee
Eq.~(\ref{eq21}) can be inverted to obtain the electrostatic two-point correlation functions in Eq.~(\ref{eq13}) in the form
\bea
\label{eq22}
&&G_\gamma(\br,\br')-v_\gamma(\br,\br')\\
&&=i\sum_{i=1}^p\lambda_iq_i \int\mathrm{d}^3\br_1v_\gamma(\br,\br_1)\lan\hk_i(\br_1)\psi_\gamma(\br')\ran.\nonumber
\eea

\subsection{Electroneutrality conditions}

\subsubsection{Global electroneutrality condition}

We derive here the global electroneutrality condition. First, we note that according to Eqs.~(\ref{eq1}) and~(\ref{eq6}), the average ion density is given by $n_i=-\delta \ln Z\G/\delta w_i(\br)$ or
\be\label{eq23}
n_i=\lambda_i\lan\hk_i(\br)\ran.
\ee
Thus, inserting the Hamiltonian~(\ref{eq11}) into the first SD identity in Eq.~(\ref{eq20}), one obtains for $\gamma={\rm l}$
\be
\label{eq24}
\int\mathrm{d}^3\br'v_{\rm l}^{-1}(\br,\br')\bar\psi_{\rm l}(\br')-\sum_{i=1}^pq_in_i=0,
\ee
with the average potential defined as $\bar\psi_{\rm l}(\br)=-i\lan\psi_{\rm l}(\br)\ran$. Taking now into account the uniformity of this potential in the bulk liquid, i.e. $\bar\psi_{\rm l}(\br)=\bar\psi_{\rm l}$, Eq.~(\ref{eq24}) reduces to
\be
\label{eq25}
\bar\psi_{\rm l}\tv_{\rm l}^{-1}(0)=\sum_{i=1}^pq_in_i.
\ee
As the infrared (IR) limit of the inverse long-range potential~(\ref{vl3}) vanishes, i.e. $\tv_{\rm l}^{-1}(q\to0)=0$, Eq.~(\ref{eq25}) finally yields the global electroneutrality condition
\be
\label{eq26}
\sum_{i=1}^pn_iq_i=0.
\ee

\subsubsection{Local electroneutrality condition}

The derivation of the local electroneutrality condition around a central charge $q_i$ located at $\br=\br\ce$ requires the calculation of  the pair distribution function between the ions of the species $i$ and $j$,
\be\label{eq44}
g_{ij}(\br,\br')=\lan\sum_{k=1}^{N_i}\sum_{l=1}^{N_j}\delta(\br-\br_{ik})\delta(\br'-\br_{jl})\frac{1-\delta_{ij}\delta_{kl}}{n_in_j}\ran_{\rm G},
\ee
where the bracket $\lan\cdot\ran_{\rm G}$ denotes the grand-canonical average, and the Kronecker deltas $\delta_{ij}$ subtract the self-interactions. Using  Eqs.~(\ref{eq1}) and~(\ref{eq6}), the identity~(\ref{eq44}) can be recast in terms of the partition function as
\be
\label{eq45}
\hspace{-2mm}g_{ij}(\br,\br')=\frac{1}{n_in_j}\frac{1}{\beta Z\G}\left\{\frac{\delta}{\delta w_j(\br')}+\delta_{ij}\delta(\br-\br')\right\}\frac{\delta Z\G}{\delta w_i(\br)}.
\ee
Substituting the functional integral form of the partition function~(\ref{eq10}) into Eq.~(\ref{eq45}), the latter can be finally expressed in terms of the functional average~(\ref{eq15}) as
\be
\label{eq46}
g_{ij}(\br,\br')=\frac{\lambda_i\lambda_j}{n_in_j}\lan \hk_i(\br)\hk_j(\br')\ran.
\ee

In the general SD identity~(\ref{eq19}) for $\gamma={\rm l}$, we set now $F[\Psi]=\lambda_i\hat{k}_i(\br\ce)$ to obtain
\bea\label{ido}
&&iq_in_i\delta^3(\br\ce-\br)=\lambda_i\int\mathrm{d}^3\br_1v_{\rm l}^{-1}(\br,\br_1)\lan\psi_{\rm l}(\br_1)\hat{k}_i(\br\ce)\ran\nonumber\\
&&\hspace{2.7cm}-in_i\sum_{j=1}^pq_jn_jH_{ij}(\br\ce,\br),
\eea
where we made use of Eqs.~(\ref{eq23}), (\ref{eq26}), and~(\ref{eq46}), and the definition of the total correlation function,
\be\label{tot}
H_{ij}(\br,\br')=g_{ij}(\br,\br')-1.
\ee
Next, integrating Eq.~(\ref{ido}) over $\br$, and accounting for the vanishing IR limit of the inverse of the potential~(\ref{vl3}), one gets the zeroth moment condition~\cite{AttardRev,Lov}
\be\label{ido1}
q_i+\sum_{j=1}^pn_jq_j\int\mathrm{d}^3\br H_{ij}(\br\ce,\br)=0.
\ee
Finally, defining the net charge density
\be
\label{ido2}
Q_i(\br\ce,\br)=q_i\delta^3(\br\ce-\br)+\sum_{j=1}^pn_jq_jH_{ij}(\br\ce,\br)
\ee
associated with the central charge $q_i$ surrounded by its spherical ion cloud of radius $R=||\br-\br\ce||$, Eq~(\ref{ido1}) can be expressed as the local electroneutrality constraint
\be
\label{locel}
\int\mathrm{d}^3\br\;Q_i(\br\ce,\br)=0.
\ee

\subsection{${\rm 2}^{\rm nd}$ moment condition and screening parameter}

\subsubsection{Two-point correlation function}

The derivation of the thermodynamic constraints and functions calculated in the remainder will be based on the net correlation function corresponding to the electrostatic interaction potential between the test charges $C_n$ and $C_m$ immersed into the electrolyte,
\be\label{cor1}
\mathcal{G}(\bR_n,\bR_m)=\left\{\frac{\partial^2\left(\beta\Omega\G\right)}{\partial C_n\partial C_m}-\frac{\partial\left(\beta\Omega\G\right)}{\partial C_n}\frac{\partial\left(\beta\Omega\G\right)}{\partial C_m}\right\}_{C_k=0}.
\ee
Inserting into Eq.~(\ref{cor1}) the dimensionless grand potential $\beta\Omega\G=-\ln Z\G$ together with the functional integral representation~(\ref{eq10}) of the partition function, one obtains
\be\label{cor2}
\mathcal{G}(\br,\br')=\sum_{\gamma=\{{\rm s,l}\}}G_\gamma(\br,\br')+\lan\psi\s(\br)\psi\lo(\br')\ran+\lan\psi\lo(\br)\psi\s(\br')\ran.
\ee 

The component $G_\gamma(\br,\br')$ of the correlation function~(\ref{cor2}) is given by Eq.~(\ref{eq22}). In order to calculate the remaining cross terms, in the general SD identity~(\ref{eq19}), we set separately $F=\psi\lo(\br')$ for $\gamma={\rm s}$,  and $F=\psi\s(\br')$ for $\gamma={\rm l}$, and use the identity~(\ref{inv}). This yields
\be
\label{cor3}
\hspace{-1mm}\lan\psi_\gamma(\br)\psi_{\gamma'}(\br')\ran=i\sum_{i=1}^p\lambda_iq_i\hspace{-1mm}\int\hspace{-1.mm}\mathrm{d}^3\br_1\hspace{-.5mm}\lan\psi_\gamma(\br)\hk_i(\br_1)\ran\hspace{-.5mm} v_{\gamma'}(\br_1,\br')
\ee
for $\gamma={\rm s}$ and $\gamma'={\rm l}$, or $\gamma={\rm l}$ and $\gamma'={\rm s}$. Plugging Eq.~(\ref{cor3}) into Eq.~(\ref{cor2}), and using Eq.~(\ref{eq8}), one obtains
\bea
\label{cor4}
\mathcal{G}(\br,\br')\hspace{-1.5mm}&=&\hspace{-1.5mm}v\ce(\br,\br')\\
&&+i\hspace{-2mm}\sum_{\gamma=\{{\rm s,l}\}}\sum_{i=1}^p\lambda_iq_i\hspace{-1mm}\int\hspace{-1mm}\mathrm{d}^3\br_1v\ce(\br,\br_1)\lan\hk_i(\br_1)\psi_\gamma(\br')\ran.\nonumber
\eea

Next, in Eq.~(\ref{eq19}), we set $F=\lambda_i\hk_i(\br_1)$, and use Eq.~(\ref{inv}) together with the net charge density~(\ref{ido2}) to get
\be\label{cor5}
\lambda_i\lan\hk_i(\br_1)\psi_\gamma(\br')\ran=in_i\int\mathrm{d}^3\br_2 Q_i(\br_1,\br_2)v_\gamma(\br_2,\br').
\ee
Substituting Eq.~(\ref{cor5}) into Eq.~(\ref{cor4}), one finally obtains an Ornstein-Zernike-like identity relating the two-point correlation function to the total correlation function,
\bea
\label{cor6}
\mathcal{G}(\br,\br')\hspace{-1mm}&=&\hspace{-1mm}v\ce(\br,\br')\\
&&\hspace{-1mm}-\hspace{-1mm}\sum_{i=1}^pn_iq_i\hspace{-1.5mm}\int\hspace{-1mm}\mathrm{d}^3\br_1\mathrm{d}^3\br_2v\ce(\br,\br_1)Q_i(\br_1,\br_2)v\ce(\br_2,\br').\nonumber
\eea

\subsubsection{Second moment condition}

In order to derive the second moment condition, we will exploit the electrical conductivity of the electrolyte originally used by Stillinger and Lovett~\cite{Lov}.  To this aim, we evaluate first the FT of Eq.~(\ref{cor6}),
\be\label{cor7}
\tgg(q)=\tv\ce(q)\left\{1-\sum_{i=1}^pn_iq_i\tq_i(q)\tv\ce(q)\right\},
\ee
where the FT of the charge density~(\ref{ido2}) reads
\be
\label{cor8}
\tq_i(q)=q_i+\sum_{j=1}^pn_jq_j\int\mathrm{d}^3\br H_{ij}(r)\frac{\sin(qr)}{qr}.
\ee 
Defining now the electric susceptibility of the electrolyte via the identity $\tgg^{-1}(q)=:\tv\ce^{-1}(q)-\tchi(q)$, Eq.~(\ref{cor7}) yields the FT of the susceptibility in the form
\be
\label{cor9}
\tchi(q)=\frac{\sum_{i=1}^pn_iq_i\tq_i(q)}{\tv\ce(q)\sum_{i=1}^pn_iq_i\tq_i(q)-1}.
\ee
Hence, the dielectric spectrum defined by the equality $\tgg(q)=:v\ce(q)/\tep(q)$ follows in terms of the susceptibility~(\ref{cor9}) as $\tep(q)=1-\tv\ce(q)\tchi(q)$, or 
\be
\label{cor10}
\tep(q)=\left\{1-\tv\ce(q)\sum_{i=1}^pn_iq_i\tq_i(q)\right\}^{-1}.
\ee

The electrical conductivity of the electrolyte implies the divergence of the dielectric spectrum~(\ref{cor10}) in the IR limit $q\to0$. In order to identify the resulting constraint, we carry out the Taylor expansion of Eq.~(\ref{cor10}) to find
\be\label{cor11}
\tep(q)=\left\{1+\frac{2\pi\ell_{\rm B}}{3}\sum_{i=1}^pn_iq_i\left[I_i^{(2)}-\frac{I_i^{(4)}}{20}q^2\right]+O\left(q^4\right)\right\}^{-1},
\ee
with the auxiliary integrals defined as
\be
\label{cor12}
I_i^{(n)}=\sum_{j=1}^pn_jq_j\int\mathrm{d}^3\br r^nH_{ij}(r).
\ee
According to Eq.~(\ref{cor11}), the IR divergence of the dielectric spectrum requires the constant term to vanish, i.e.
\be
\label{cor13}
\frac{2\pi\ell_{\rm B}}{3}\sum_{i,j}n_in_jq_iq_j\int\mathrm{d}^3\br\;r^2H_{ij}(r)=-1,
\ee
which corresponds precisely to the second moment condition originally derived by Stillinger and Lovett~\cite{Lov,AttardRev}.

\subsubsection{Dressed ion charge}

Within the present formalism, we recover here Kjellander's dressed ion charge~\cite{Kj1}. To this aim, we express first the potential $\phi_i(\br)$ induced by a source charge $q_i$,
\be\label{cor14}
\phi_i(\br)=\int\mathrm{d}^3\br'v\ce(\br,\br')Q_i(\br')=:\int\mathrm{d}^3\br'\mathcal{G}(\br,\br')\rho^*_i(\br').
\ee
In Eq.~(\ref{cor14}), the second equality defining the dressed charge density $\rho^*_i(\br)$ associated with the central charge $q_i$ and its close counterion cloud implies that the interaction of the dressed charges is mediated by the net two-point correlation function dressed by the far-cloud screening. 

Next, we evaluate the FT of Eq.~(\ref{cor14}) to obtain the dressed charge density in reciprocal space as
\be
\label{cor15}
\tilde{\rho}_i^*(q)=\tilde{Q}_i(q)\tep(q),
\ee
where we used the dielectric spectrum~(\ref{cor10}). The net dressed charge follows from the integral of the dressed charge density over the entire space, i.e. $q_i^*=\int\mathrm{d}^3\br\rho^*_i(\br)=\tilde{\rho}_i^*(0)$. Thus, evaluating the IR limit of Eq.~(\ref{cor15}), and taking into account the moment conditions~(\ref{ido1}) and~(\ref{cor13}), one finally obtains the dressed charge in terms of the integrals in Eq.~(\ref{cor12}) as follows,
\be
\label{cor16}
q_i^*=\frac{5I_i^{(2)}}{\pi\ell_{B}\sum_{j=1}^pn_jq_jI_j^{(4)}}.
\ee

\subsubsection{Screening parameter}

In order to derive the screening parameter, we combine first Eqs.~(\ref{cor7}) and~(\ref{cor9}) to express the correlation function in real space as
\be
\label{cor17}
\mathcal{G}(r)=\int\frac{{\rm d}^3\br}{\left(2\pi\right)^3}\frac{4\pi\ell_{\rm B}}{q^2-4\pi\ell_{\rm B}\tchi(q)}e^{i\bq\cdot\br}.
\ee
According to the asymptotic analysis of the OZ equation, at large separation distances, the correlation function~(\ref{cor17}) exhibits an exponential decay~\cite{AttardRev},  i.e.
\be
\label{cor18}
\mathcal{G}\left(r\gg\kappa^{-1}\right)\approx\frac{\ell_{\rm B}}{r}e^{-\kappa r}=\int\frac{{\rm d}^3\br}{\left(2\pi\right)^3}\frac{4\pi\ell_{\rm B}}{q^2+\kappa^2}e^{i\bq\cdot\br}.
\ee
The comparison of the integrals in Eqs.~(\ref{cor17})-(\ref{cor18}) indicates the existence of poles fulfilling the condition $\left[q^2-4\pi\ell_{\rm B}\tchi(q)\right]_{q=i\kappa}=0$. This yields the following non-linear equation satisfied by the screening parameter
\be
\label{cor19}
\kappa^2=-4\pi\ell_{\rm B}\tchi(i\kappa)
\ee
originally derived by Kjellander~\cite{Kj1995}.

In the present work, we limit ourselves to the leading order perturbative solution of the identity~(\ref{cor19}). Thus, Taylor-expanding Eqs.~(\ref{cor9}) and~(\ref{cor19}) in terms of the parameter $\kappa$, and accounting for the moment conditions~(\ref{ido1}) and~(\ref{cor13}), the screening parameter follows as a modified DH identity involving the dressed charge~(\ref{cor16}),
\be
\label{cor20}
\kappa^2\approx4\pi\ell_{\rm B}\sum_{i=1}^pn_iq_iq^*_i.
\ee

\subsection{Generalized cumulant approach}
\label{gencum}

The equations preceding Eq.~(\ref{cor20}) are exact identities. However, the non-linearity of the Hamiltonian functional~(\ref{eq11}) does not allow the exact calculation of the functional averages in Eqs.~(\ref{eq22}) and~(\ref{eq46}). Thus,  in the present work, these averages will be evaluated via a generalized version of our cumulant expansion scheme introduced in Ref.~\cite{Buyuk2024}.

Our approach is based on the exact splitting of the Hamiltonian~(\ref{eq15}) into a Gaussian reference component $H_0[\ph]$ whose form will be specified below, and a second component $\delta H[\ph]$ including the non-linearities beyond Gaussian-level that will be treated perturbatively, i.e.
\be\label{eq27}
H[\ph]=H_0[\ph]+t\delta H[\ph].
\ee
In Eq.~(\ref{eq27}), the expansion parameter $t$ of unit magnitude will enable us to keep track of the perturbative order. 

In order to simplify the notation, in the remainder, the arguments of the functionals will be omitted. Plugging now the decomposition~(\ref{eq27}) into Eq.~(\ref{eq15}), and Taylor-expanding the result at the first perturbative order $O(t)$, the average of the general functional $F$ follows as
\be
\label{eq28}
\lan F\ran=\lan F\ran_0-t\left[\lan\beta\delta HF\ran_0-\lan\beta\delta H\ran_0\lan F\ran_0\right]+O\left(t^2\right).
\ee
In Eq.~(\ref{eq28}), the Gaussian-level averages and the corresponding partition function have been defined in the form
\be
\label{eq29}
\lan F\ran_0=\frac{1}{Z_0}\int\mathcal{D}\ph\;e^{-\beta H_0}F;\hspace{5mm}Z_0=\int\mathcal{D}\ph\;e^{-\beta H_0}.
\ee

Within the present formalism, the functional averages over the fluctuating potentials $\psi_\gamma(\br)$ associated with the short-range ($\gamma=\{{\rm h,s}\}$) and the long-range ($\gamma={\rm l}$) interactions will be evaluated via the virial and the cumulant treatment, respectively. To this aim, the Gaussian reference Hamiltonian in Eq.~(\ref{eq27}) is chosen in the form
\bea
\label{eq30}
\beta H_0&=&\int\frac{\mathrm{d}^3\br\mathrm{d}^3\br'}{2}\left\{\sum_{\gamma=\{{\rm h,s}\}}\psi_\gamma(\br)v^{-1}_\gamma(\br-\br')\psi_\gamma(\br')\right.\nonumber\\
&&\left.\hspace{2.2cm}+\psi_{\rm l}(\br)G_{\rm l}^{-1}(\br-\br')\psi_{\rm l}(\br')\right\},
\eea
where the variances of the short-range fluctuations ($\gamma=\{{\rm h,s}\}$) are incorporated via the bare interaction potentials $v_\gamma(\br,\br')$, and the long-range fluctuations are taken into account with the two-point correlation function $G_{\rm l}(\br-\br')$ that will be obtained from the solution of the SD identity~(\ref{eq21}) for $\gamma={\rm l}$. Finally, due to the specific form of the reference Hamiltonian in Eq.~(\ref{eq30}), the non-linear Hamiltonian component in Eq.~(\ref{eq27}) to be treated perturbatively follows from Eq.~(\ref{eq11}) as
\bea
\label{eq31}
\beta\delta H&=&\int\frac{\mathrm{d}^3\br\mathrm{d}^3\br'}{2}\psi_{\rm l}(\br)\left[v_{\rm l}^{-1}(\br,\br')-G_{\rm l}^{-1}(\br,\br')\right]\psi_{\rm l}(\br')\nonumber\\
&&-\sum_{i=1}^p\lambda_i\int\mathrm{d}^3\br\;\hk_i(\br).
\eea

\subsubsection{Cumulant-level calculation of the ion fugacities}

We relate here the ion fugacities to the bulk salt concentrations. To this aim, we carry out the cumulant expansion of the functional average in Eq.~(\ref{eq23}) according to Eq.~(\ref{eq28}), calculate the Gaussian averages defined in Eq.~(\ref{eq29}), and plug into the resulting expression the expansion of the long-wavelength kernel into its Gaussian-level and first order cumulant components,
\be
\label{eq32}
G_{\rm l}(\br-\br')=G_{{\rm l},0}(\br-\br')+t\;G_{{\rm l},1}(\br-\br')+O\left(t^2\right).
\ee
After lengthy algebra, one obtains at the order $O\left(t\right)$
\bea
\label{eq33}
n_i&=&\Lambda_i-t\Lambda_i\left\{\frac{q_i^2}{2}G_{{\rm l},1}(0)-\sum_{j=1}^p\Lambda_j\int\mathrm{d}^3\br\; h_{ij}(\br)\right\}\nonumber\\
&&-t\Lambda_iq_i^2\int\frac{\mathrm{d}^3\br_1\mathrm{d}^3\br_2}{2}\left[G_{{\rm l},0}^{-1}-v_{\rm {\SB l}}^{-1}\right]_{\br_1,\br_2}\\
&&\hspace{2.8cm}\times G_{{\rm l},0}(\br-\br_1)G_{{\rm l},0}(\br-\br_2),\nonumber
\eea
with the rescaled fugacity $\Lambda_i=\lambda_i\;e^{-q_i^2\delta G_{{\rm l},0}(0)/2}$ \SB{and the Mayer function defined as}
\be
\label{eq34II}
h_{ij}(\br)=e^{-v\h(\br)-q_iq_j\left[v_{\rm s}(\br)+G_{{\rm l},0}(\br)\right]}-1,
\ee
and the self-energy $\delta G_{{\rm l},0}(0)=[G_{{\rm l},0}(\br,\br')-v\ce(\br,\br')]_{\br'\to\br}$. Finally, in order to invert Eq.~(\ref{eq33}), we insert into the latter the expansion $\Lambda_i=\Lambda_i^{(0)}+t\Lambda_i^{(1)}+O\left(t^2\right)$ and identify the terms of different perturbative orders. This yields the ion fugacity as a function of concentration as
\bea\label{eq35}
\Lambda_i&=&n_i+t\frac{q_i^2}{2}n_iG_{{\rm l},1}(0)-tn_i\sum_{j=1}^pn_j\int\mathrm{d}^3\br\; h_{ij}(\br)\nonumber\\
&&+tq_i^2n_i\int\frac{\mathrm{d}^3\br_1\mathrm{d}^3\br_2}{2}\left[G_{{\rm l},0}^{-1}-v_{\rm {\SB l}}^{-1}\right]_{\br_1,\br_2}\\
&&\hspace{3.0cm}\times G_{{\rm l},0}(\br-\br_1)G_{{\rm l},0}(\br-\br_2).\nonumber
\eea

\subsubsection{Cumulant expansion of the variational Eq.~(\ref{eq16})}

For the cumulant-level calculation of the variational identity~(\ref{eq16}), we evaluate first the Green's function associated with the long-range ion interactions. To this aim, we carry out the cumulant expansion~(\ref{eq28}) of the SD identity~(\ref{eq21}) for $\gamma={\rm l}$, and calculate the functional averages defined by Eq.~(\ref{eq29}). Inserting into the resulting expression the kernel expansion~(\ref{eq32}) together with the fugacity identity~(\ref{eq35}), the equations satisfied by the kernel components of different orders follow in the form
\bea
\label{eq36}
\hspace{-5mm}&&\int\mathrm{d}^3\br_1v_{\rm l}^{-1}(\br-\br_1)G_{{\rm l},0}(\br_1-\br')+\sum_{i=1}^pn_iq_i^2G_{{\rm l},0}(\br-\br')\nonumber\\
\hspace{-5mm}&&=\delta^3(\br-\br');\\
\label{eq37}
\hspace{-5mm}&&\int\mathrm{d}^3\br_1v_{\rm l}^{-1}(\br-\br_1)G_{{\rm l},1}(\br_1-\br')+\sum_{i=1}^pn_iq_i^2G_{{\rm l},1}(\br-\br')\nonumber\\\
\hspace{-5mm}&&=-\sum_{i,j}n_in_jq_iq_j\int\mathrm{d}^3\br_1\left\{h_{ij}(\br-\br_1)+q_iq_jG_{{\rm l},0}(\br-\br_1)\right\}\nonumber\\
\hspace{-5mm}&&\hspace{3.5cm}\times G_{{\rm l},0}(\br_1-\br').
\eea
Solving now the kernel identities~(\ref{eq36})-(\ref{eq37}) in Fourier space, the Fourier-transformed components of the long-range Green's function follow as
\bea
\label{eq38}
\tG_{{\rm l},0}(q)&=&\left[\tv_{\rm l}^{-1}(q)+\sum_in_iq_i^2\right]^{-1};\\
\label{eq39}
\tG_{{\rm l},1}(q)&=&-\tG_{{\rm l},0}^2(q)\sum_{i,j}n_in_jq_iq_j\left[\may+q_iq_j\tG_{{\rm l},0}(q)\right].\nonumber\\
\eea
In Eq.~(\ref{eq39}), the FT of the Mayer function reads
\bea\label{eq40}
\may\hspace{-2mm}&=&\hspace{-2mm}-\frac{4\pi}{q^3}\left[\sin(qd)-qd\cos(qd)\right]\\
&&\hspace{-2mm}+4\pi\int_d^\infty\mathrm{d}rr^2\frac{\sin(qr)}{qr}\left\{e^{-q_iq_j\left[v\s(r)+G_{{\rm l},0}(r)\right]}-1\right\}.\nonumber
\eea

Adding now Eqs.~(\ref{eq38})-(\ref{eq39}), at the order $O\left(t\right)$, the FT of the long-wavelength kernel~(\ref{eq32}) becomes 
\bea
\label{eq41}
\tG_{\rm l}(q)&=&\tv_{\rm l}(q)-\sum_in_iq_i^2\tG_{{\rm l},0}(q)\tv_{\rm l}(q)\\
&&-t\tG_{{\rm l},0}^2(q)\sum_{i,j}n_in_jq_iq_j\left[\may+q_iq_j\tG_{{\rm l},0}(q)\right].\nonumber
\eea
Then, in order to calculate the short-wavelength propagator, we carry out the cumulant expansion~(\ref{eq28}) of the functional average on the r.h.s. of Eq.~(\ref{eq22}) for $\gamma={\rm s}$, evaluate the Gaussian averages according to Eq.~(\ref{eq29}), and plug into the result the kernel expansion~(\ref{eq32}) and the fugacity identity~(\ref{eq35}). Passing to the reciprocal Fourier space, one gets at the order $O\left(t\right)$
\be
\label{eq42}
\tG_{\rm s}(q)=\tv_{\rm s}(q)-\sum_in_iq_i^2\tv^2_{\rm s}(q)-t\sum_{i,j}n_in_jq_iq_j\may\tv^2_{\rm s}(q).
\ee
Finally, plugging the kernels~(\ref{eq41})-(\ref{eq42}) into the variational Eq.~(\ref{eq16}), at the order $O\left(t\right)$, one obtains
\bea
\label{eq43}
&&\sum_{i,j}n_in_jq_iq_j\int_0^\infty\mathrm{d}qq^2\left[\may+q_iq_j\tG_{{\rm l},0}(q)\right]\nonumber\\
&&\hspace{3cm}\times\left[1-\tG_{{\rm l},0}^2(q)\tv_{\rm l}^{-2}(q)\right]\partial_\sigma\tv_{\rm l}(q)\nonumber\\
&&=0.
\eea

In the present work, the splitting parameter $\sigma$ was obtained from the numerical solution of the variational identity~(\ref{eq43}) via a dichotomy algorithm. We also note that as the Fourier-transformed kernel~(\ref{eq38}) associated with the long-range potential~(\ref{vl1}) cannot be inverted analytically, the real-space kernel $G_{{\rm l},0}(r)$ in the exponential term of Eq.~(\ref{eq40}) should be computed numerically.

\subsubsection{Cumulant-level evaluation of the correlation functions}

The evaluation of the thermodynamic functions investigated in this work requires the explicit calculation of the pair distribution function~(\ref{eq46}). To this aim, we carry out the cumulant expansion~(\ref{eq28}) of Eq.~(\ref{eq46}), evaluate the functional averages according to Eq.~(\ref{eq29}), and incorporate the kernel expansion~(\ref{eq32}) together with the fugacity identity~(\ref{eq35}). After lengthy algebra, the total correlation function defined by Eq.~(\ref{tot}) takes the form
\be
\label{eq47}
H_{ij}(\br-\br')=h_{ij}(\br-\br')+t\;\left[h_{ij}(\br-\br')+1\right]T_{ij}(\br-\br').
\ee
In Eq.~(\ref{eq47}), we defined the auxiliary function
\bea\label{eq48}	
\hspace{-4mm}&&T_{ij}(\br-\br')=\sum_{k=1}^pn_k\int\mathrm{d}^3\br_1\left\{h_{ik}(\br-\br_1)h_{kj}(\br_1-\br')\right.\\
\hspace{-4mm}&&\hspace{3.5cm}\left.-q_iq_jq_k^2G_{{\rm l},0}(\br-\br_1)G_{{\rm l},0}(\br_1-\br')\right\}\nonumber\\
\hspace{-4mm}&&\hspace{2cm}-q_iq_jG_{{\rm l},1}(\br-\br').\nonumber
\eea
Passing to the reciprocal space, Eq.~(\ref{eq48}) can be expressed in terms of the Fourier-transformed functions~(\ref{eq38})-(\ref{eq40}) as
\bea
\label{eq49}
T_{ij}(r)&=&\int_0^\infty\frac{\mathrm{d}qq^2}{2\pi^2}\frac{\sin(qr)}{qr}\\
&&\hspace{1cm}\times\left\{\sum_{k=1}^pn_k\left[\tilde{h}_{ik}(q)\tilde{h}_{kj}(q)-q_iq_jq_k^2\tG_{{\rm l},0}^2(q)\right]\right.\nonumber\\
&&\left.\hspace{1.5cm}-q_iq_j\tG_{{\rm l},1}(q)\right\}.\nonumber
\eea

\section{Results and discussion}
\label{res}

In this section, we compare the predictions of the present formalism with numerical and experimental data from the literature. For the interpretation of the results reported herein, it is essential to note that the CCDH theory of Ref.~\cite{Buyuk2024} follows from the present formalism in the limit of vanishing splitting parameter $\sigma\to0$ where the short-range potential~(\ref{vl4}) in Eq.~(\ref{eq40}) vanishes ($v\s(r)\to0$), and the Gaussian component of the long-range correlator reduces to the electrostatic WC-level DH potential, i.e. \SB{$G_{{\rm l},0}(r)\to v_{\rm DH}(r)=\ell_{\rm B}e^{-\kappa_0 r}/r$}, where $\kappa_0=\sqrt{8\pi\ell_{\rm B}n_i}$ is the DH screening parameter. Thus, the CCDH theory corresponds to the electrostatic WC limit of the SCDH approach. This implies that the pure HC interactions are embodied by the SCDH formalism at the same level of accuracy as the CCDH theory.

\SB{\subsection{Radial distribution and charge density functions}}

\begin{figure}
\includegraphics[width=1.\linewidth]{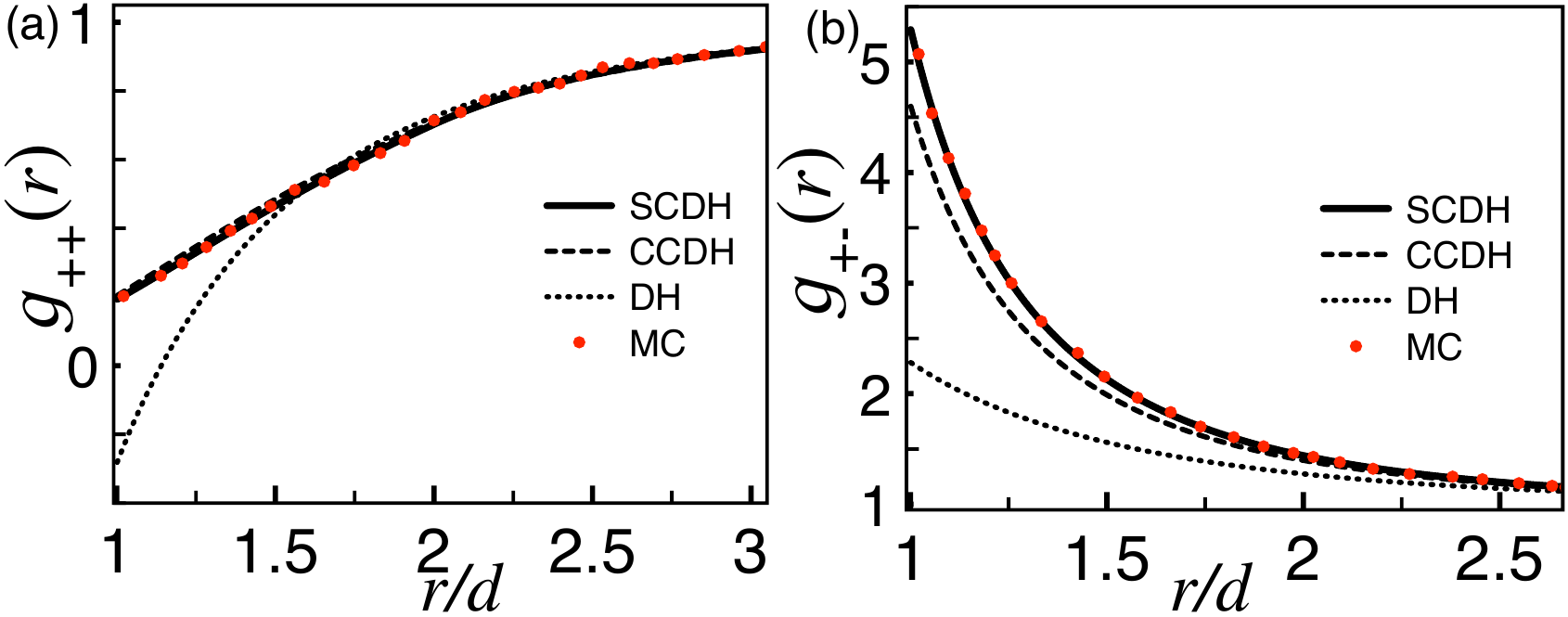}
\caption{(Color online)  \SB{Pair distribution functions associated with a central monovalent cation at the packing fraction $\eta=\pi n_id^3/3=0.01$ and the reduced ion size $d/\ell_{\rm B}=1/3$. At the temperature $T = 298$ K and the dielectric constant $\ew = 78.5$, these parameters correspond to the ion size $d\approx2.38$ {\AA} and density $n_i\approx1.17$ M. The MC data are from Fig. 1 of Ref.~\cite{NetzMC}.}} 
\label{fig1}
\end{figure}

\SB{Fig.~\ref{fig1} compares the theoretical predictions  for the pair distribution functions around a central monovalent cation with MC simulation results (red symbols) at the ion size $d\approx2.38$ {\AA} and concentration $n_i\approx1.17$ M. One sees that the cation-cation distribution function underestimated by the DH approximation $g_{ij}(r)\approx1-q_iq_jv_{\rm DH}(r)$ (dotted curve) is accurately reproduced by both the CCDH (dashed curve) and the SCDH formalism (solid curve). In the case of the cation-anion distribution function, while the CCDH approach agrees with MC data only qualitatively, the simulation result is reproduced by the SCDH formalism with quantitative accuracy.}

\begin{figure}
\includegraphics[width=1.\linewidth]{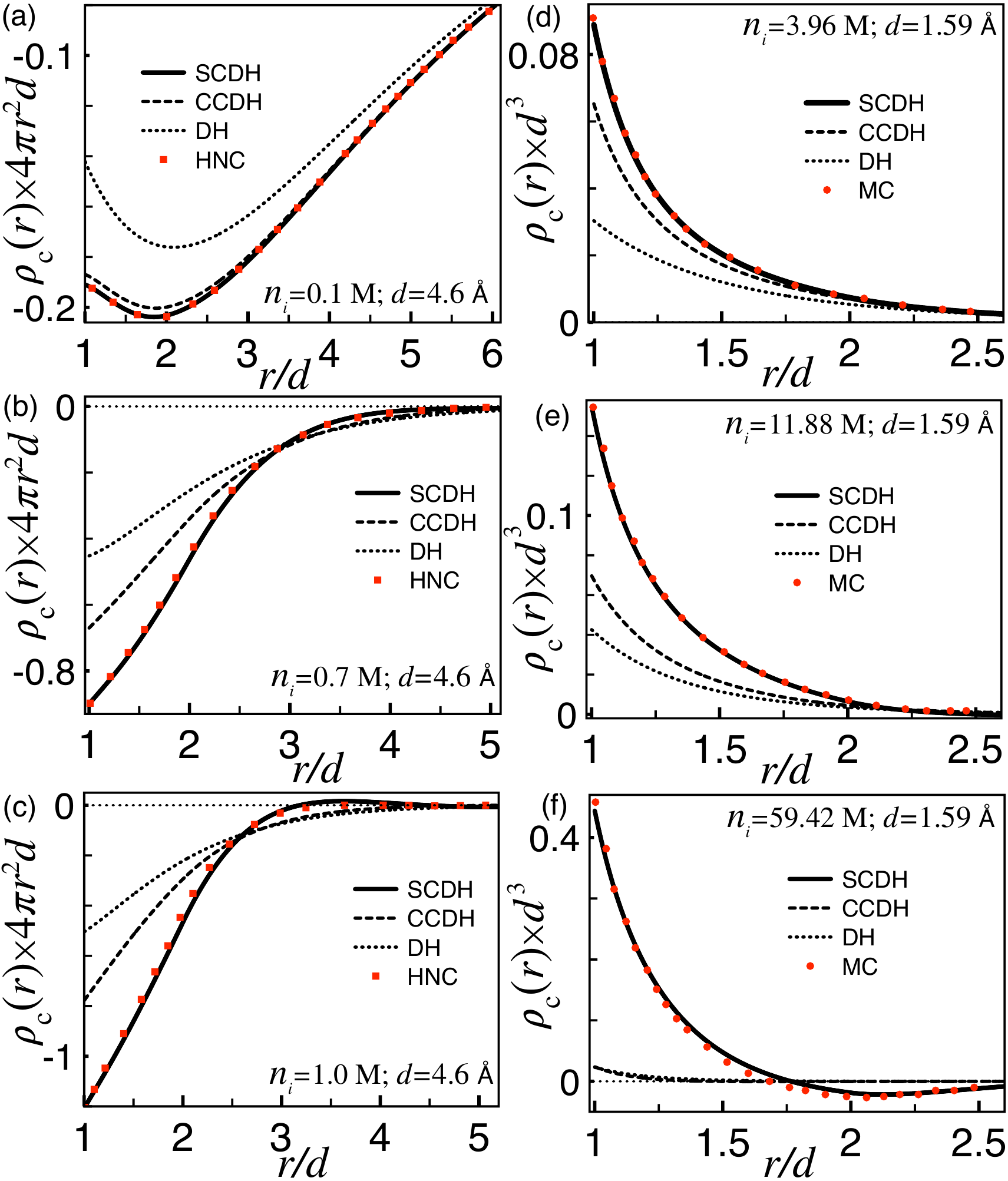}
\caption{(Color online)  \SB{Charge density $\rho\ce(r)=q_+n_+H_{i+}(r)+q_-n_-H_{i-}(r)$ around a central (a)-(c) cation ($i=+$) and (d)-(f) anion ($i=-$) in 1:1 liquids. In (a)-(c), the HNC data are from Fig. 6 of Ref.~\cite{Kj2020}. The temperature and the dielectric permittivity are $T=298$ K and $\e=78.4$. The remaining parameters are provided in the legends. In (d)-(f), the MC data are from Fig. 8.B of Ref.~\cite{NetzMC}. From top to bottom, the packing fractions are $\eta=0.01$, $0.03$, and $0.15$. The reduced ion size is $d/\ell_{\rm B}=2/9$. The corresponding physical parameters are provided for $T=298$ K and $\e=78.5$ in the legends.}} 
\label{fig2}
\end{figure}

\SB{In Figs.~\ref{fig2}(a)-(c), the theoretical charge densities associated with a central cation are compared with HNC predictions at the larger ion size $d=4.6$ {\AA} characterized by substantial HC correlations. One sees that the DH prediction ignoring the HC ion size is qualitatively inaccurate in all three plots. Then, the CCDH approach agrees qualitatively with the HNC data at the dilute concentration $n_i=0.1$ M, but significantly deviates from the HNC result at larger concentrations. Finally, the SCDH formalism agrees quantitatively with the HNC data in the entire concentration regime covered by the figure.}

\SB{Figs.~\ref{fig2}(d)-(f) display the charge distributions around a central anion at the significantly smaller HC size $d=1.59$ {\AA} and large molar concentrations. This regime is characterized by the competition between pronounced HC interactions and substantial electrostatic correlations induced by the close interionic approach. One sees that the SCDH formalism improves the accuracy of the  DH and CCDH predictions in a sizable fashion, and agrees quantitatively with the MC data over the entire concentration range considered in the figure. In particular, at the atypically large ion concentration of Fig.~\ref{fig2}(f) characterized by a weak charge inversion (CI) effect, the SCDH formalism corrects the CCDH prediction by an order of magnitude, and reproduces the CI with reasonable accuracy.}

\begin{figure}
\includegraphics[width=1.\linewidth]{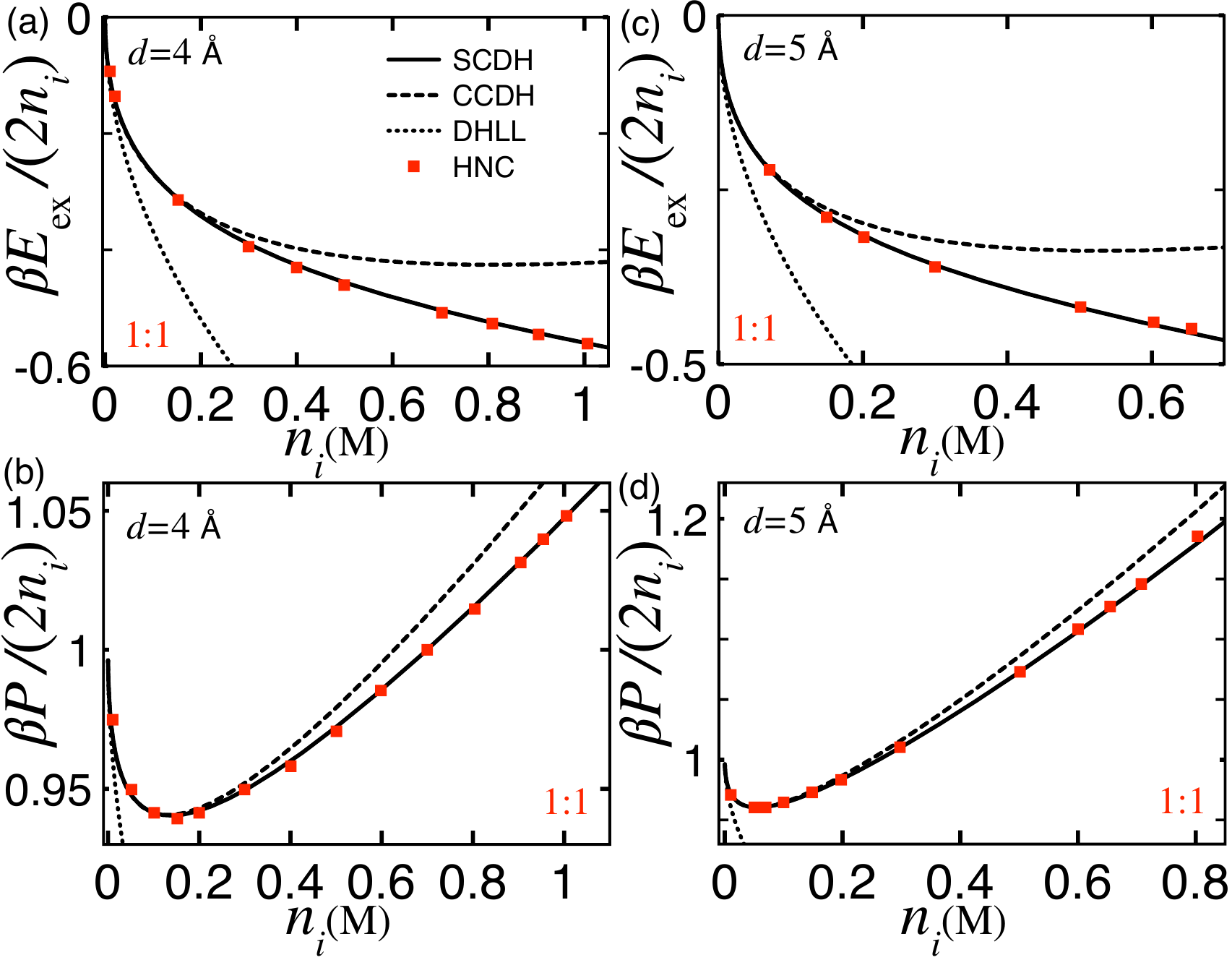}
\caption{(Color online) (a)-(c) Internal energy~(\ref{eq50}) and (b)-(d) pressure~(\ref{eq51}) of 1:1 solutions against salt concentration at two different HC sizes indicated in the legends. The temperature and the dielectric constant of the liquid are $T = 300$ K and $\ew = 78.5$. Solid curves: SCDH theory. Dashed curves: CCDH formalism of Ref.~\cite{Buyuk2024}. Dotted curves: DHLLs. Red squares: HNC data from Figs. 3 and 4 of Ref.~\cite{AttardPRE}.} 
\label{fig3}
\end{figure}

\subsection{Excess energy and pressure}

\SB{The energy density and pressure of an electrolyte mixture can be expressed in terms of the total correlation function~(\ref{eq47}) considered in the previous part as~\cite{Buyuk2024}}
\bea\label{eq50}
&&\beta E_{\rm ex}=2\pi\ell_{\rm B}\sum_{i=1}^p\sum_{j=1}^pn_in_jq_iq_j\int_d^\infty\mathrm{d}rrH_{ij}(r);\\
\label{eq51}
&&\beta P=\sum^p_{i=1} n_i+\frac{2\pi}{3}d^3\left(\sum_{i=1}^pn_i\right)^2\\
&&\hspace{9mm}+\frac{2\pi}{3}d^3\sum_{i=1}^p\sum_{k=1}^pn_in_kH_{ik}(d^+)+\frac{1}{3}\beta E_{\rm ex}.\nonumber
\eea
On the r.h.s. of Eq.~(\ref{eq51}), the second term following the ideal gas pressure is an excluded volume contribution corresponding to the first order virial expansion of the Carnahan-Starling (CS) equation of state~\cite{Buyuk2024}. Then, the third term incorporating the contact pair densities include the second order virial expansion term of the CS pressure~\cite{Buyuk2024}. Finally, these repulsive contributions originating from HC interactions are counterbalanced by the attractive internal energy (fourth term) mainly governed by the electrostatic coupling of opposite charges.

\SB{Figs.~\ref{fig3}(a)-(d) compare the excess energy~(\ref{eq50}) and pressure~(\ref{eq51}) of the SCDH approach with HNC data for 1:1 liquids at two different ion sizes. The plots also display the CCDH predictions, and the DH limiting laws (DHLLs),}
\be
\label{eq52}
\beta E_{{\rm ex,DH}}=-\frac{\kappa_0^3}{8\pi};\hspace{5mm}\beta P_{\rm DH}=\sum_{i=1}^pn_i-\frac{\kappa_0^3}{24\pi}.
\ee
\SB{In consistency with the ion distribution plots in Figs.~\ref{fig1}-\ref{fig2}, these energy and pressure plots} show that the variational scheme at the basis of the SCDH formalism significantly improves the agreement of the CCDH theory with HNC data.  Namely, while the CCDH formalism extends the validity of the DHLLs from $n_i\sim10^{-2}$ M to $n_i\sim0.2$ M, the SCDH results agree quantitatively with HNC data up to $n_i\sim1.0$ M. Hence, for monovalent ions with typical hydration radii $d\sim4$ {\AA}, the asymmetric treatment of the short- and long-range ion interactions enables the SCDH approach to accurately predict both the charge correlation-driven internal energy and the HC-dominated liquid pressure up to the molar concentration regime.

\begin{figure}
\includegraphics[width=1.\linewidth]{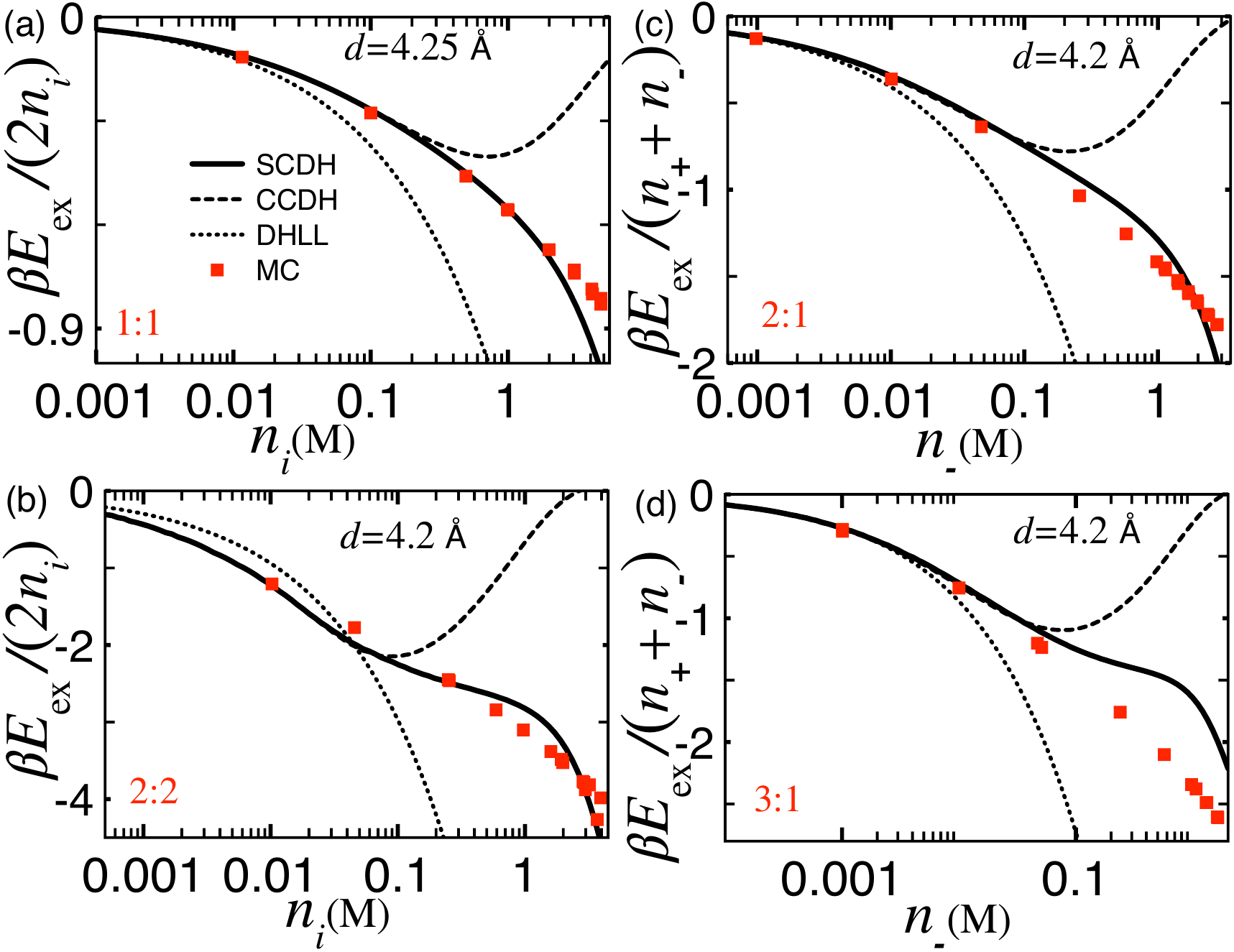}
\caption{(Color online) Internal energy~(\ref{eq50}) against salt concentration. The ion valencies and sizes are indicated in the legends. The temperature and the dielectric constant are $T = 298$ K and $\ew = 78.5$. The curves have the same signification as in Fig.~\ref{fig3}. The MC data (red squares) are from (a) Table I, (b) II, (c) III, and (d) IV of Ref.~\cite{Valleau}.} 
\label{fig4}
\end{figure}

Fig.~\ref{fig4}(a) confronts the theoretical internal energy predictions for 1:1 solutions with MC simulations over a larger concentration regime. The quantitative accuracy of the SCDH approach up to $n_i\sim1.0$ M is equally confirmed by this plot. At larger concentrations $n_i\gtrsim2.0$ M where the SCDH prediction overestimates the MC data, the qualitative agreement of the present formalism with simulations is significantly better than that of the CCDH approach exhibiting a trend reversal, and an unphysical rise of the energy previously reported in Ref.~\cite{Buyuk2024}.

Figs.~\ref{fig4}(b)-(d) show that for multivalent electrolytes characterized by stronger electrostatic coupling, the quantitative agreement of the SCDH formalism with MC data is limited to submolar concentrations. However, in the vicinity of the molar concentration regime, the SCDH approach can qualitatively reproduce the trend and magnitude of the simulation data with a considerably better accuracy than the CCDH formalism. 

\begin{figure*}
\includegraphics[width=1.\linewidth]{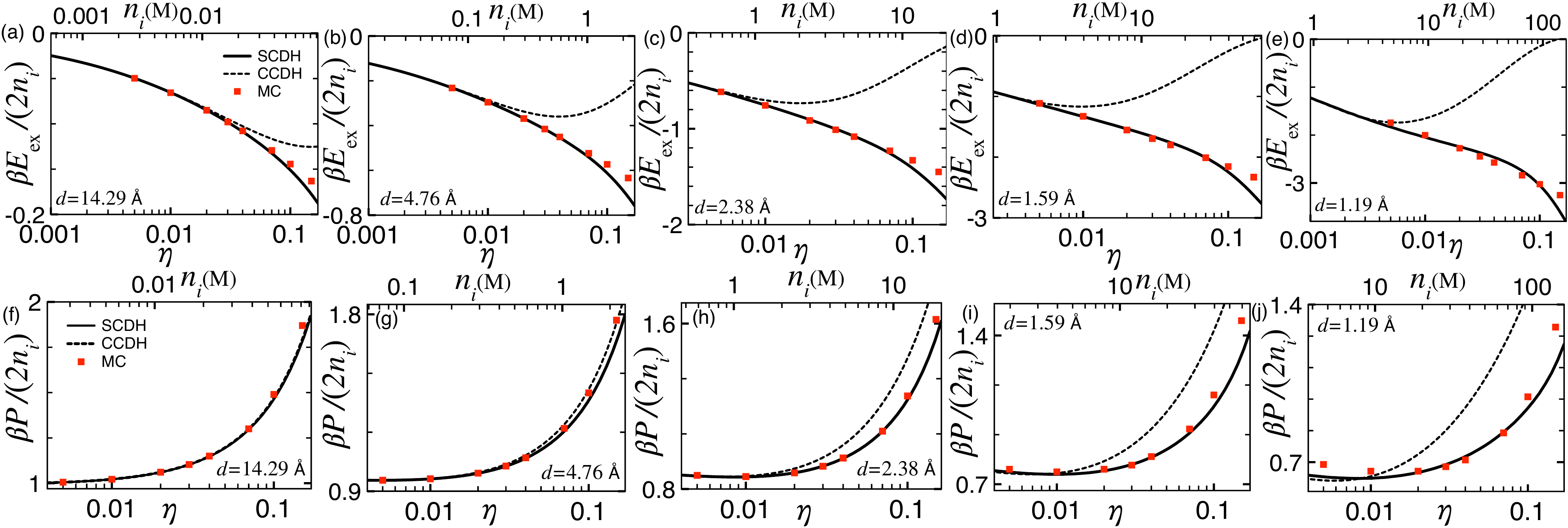}
\caption{(Color online) (a)-(e) Internal energy~(\ref{eq50}) and (f)-(j) pressure~(\ref{eq51}) of monovalent ions against their packing fraction  (lower axis) and concentration (upper axis) at the temperature $T=298$ K, the dielectric permittivity $\e_{\rm w}=78.5$, and the HC sizes indicated in the legends~\cite{rem1}. MC data are from (a)-(e) Table I and (f)-(j) Table II of Ref.~\cite{NetzMC}.} 
\label{fig5}
\end{figure*}
We investigate now the effect of HC size on the accuracy of the SCDH formalism. In Fig.~\ref{fig5}(a), the comparison \SB{of the energy curves} with MC data shows that at the particularly large solute size $d=14.29\;{\rm {\AA}}>\ell_{\rm B}\approx7.1$ {\AA} where the strong attenuation of opposite charge attraction by HC repulsion gives rise to weak electrostatic coupling conditions, the SCDH formalism brings a moderate correction to the CCDH \SB{approach}. Then, in the typical case of smaller ion sizes $d<\ell_{\rm B}$ of Figs.~\ref{fig5}(b)-(e) where the shorter interionic approach distances enhance the strength of the electrostatic correlations, the SCDH formalism substantially improves the quantitative accuracy of the CCDH approach, extending its validity limit by several factors into the molar concentration regime.

In Figs.~\ref{fig5}(f)-(j), we illustrate the effect of solute size on the liquid pressure~(\ref{eq51}). At the ion size $d=14.29$ {\AA} where the HC component of the pressure dominates its electrostatic component, the SCDH and CCDH predictions roughly equivalent to the second order virial expansion of the CS pressure~\cite{Buyuk2024} both agree with MC data up to the packing fraction $\eta\approx0.1$. However, at the solute sizes $d<\ell_{\rm B}$ of Fig.~\ref{fig5}(g)-(j) where the attractive energy component of the pressure embodying charge correlations becomes comparable with its repulsive HC component, the SCDH approach provides a more accurate prediction than the CCDH theory. In particular, at the substantially small solute size of Fig.~\ref{fig5}(i) characterized by strong electrostatic coupling conditions,  the SCDH formalism can reproduce with reasonable accuracy the plateau of the osmotic coefficient driven by the pronounced competition between opposite charge attraction and core repulsion over a broad concentration range.

\subsection{Comparison of different splitting schemes}
\label{comp}

The systematic comparison of the present formalism with simulations involving different salt valencies, concentrations, and HC sizes showed that the splitting of the short- and long-range interactions enables the SCDH formalism to handle electrostatic correlations in a significantly more accurate fashion than the CCDH approach. Here, we compare the underlying splitting scheme associated with the filter operator of fourth differential order~(\ref{vl1}) with the approach involving the inverse kernel  $v^{-1}\lo(\br,\br')=\left(1-\sigma^2\nabla^2\right)v\ce^{-1}(\br,\br')$ including the filter operator of second differential order employed by Santangelo~\cite{Santangelo}, and the splitting approach of Ref.~\cite{sp} using error functions. Together with the constraint~(\ref{eq8}), these two different choices yield respectively the following short-range potentials,
\bea
\label{eq53}
\hspace{-5mm}\tv_{\rm l}(q)&=&\frac{4\pi\ell_{\rm B}}{q^2}\left(1+\sigma^2 q^2\right)^{-1}\;\Rightarrow\;v_{\rm s}(r)=\frac{\ell_{\rm B}}{r}e^{-r/\sigma};\\
\label{eq54}
\hspace{-5mm}\tv_{\rm l}(q)&=&\frac{4\pi\ell_{\rm B}}{q^2}e^{-\sigma^2q^2}\;\Rightarrow\;v_{\rm s}(r)=\frac{\ell_{\rm B}}{r}{\rm erfc}\left[\frac{r}{2\sigma}\right],
\eea
where ${\rm erfc}(x)$ is the complementary error function~\cite{math}. 

\begin{figure}
\includegraphics[width=1.\linewidth]{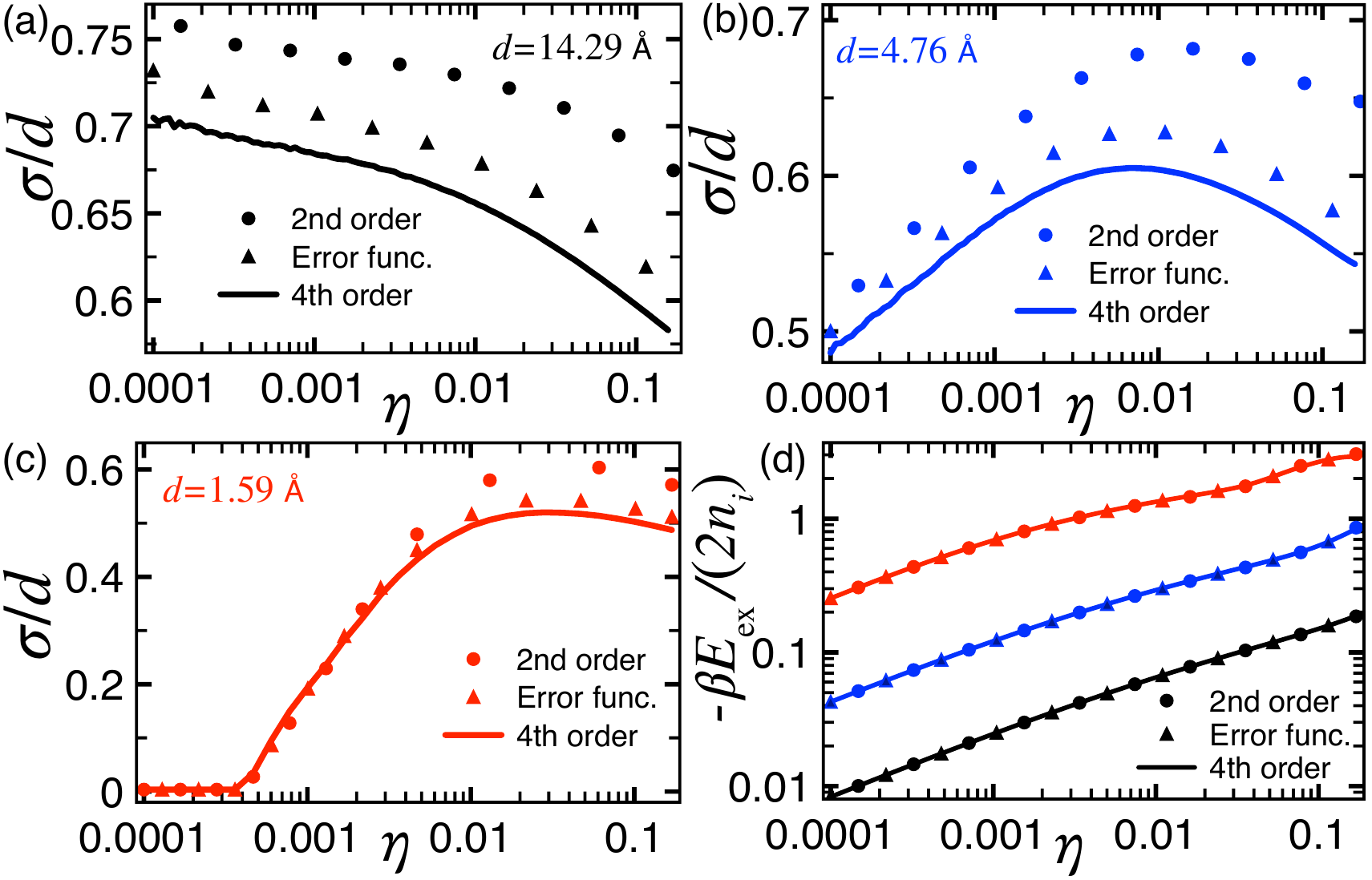}
\caption{(Color online) (a)-(c) Splitting parameter solving Eq.~(\ref{eq43}) rescaled by the HC size against packing fraction at the model parameters of Fig.~\ref{fig5}. (d) Minus the excess energy~(\ref{eq50}) at the ion sizes corresponding to the colors in (a)-(c). In all plots, the solid curves, the disks, and the triangles correspond to the solution of the variational Eq.~(\ref{eq43}) with the short-range potentials~(\ref{vl4}), (\ref{eq53}), and~(\ref{eq54}), respectively.} 
\label{fig6}
\end{figure}

In Figs.~\ref{fig6}(a)-(c), we show that the splitting lengths $\sigma$ obtained from the solution of Eq.~(\ref{eq43}) with these three distinct splitting potentials are significantly close, and they exhibit qualitatively similar behavior with the variation of the packing fraction or equivalently the salt concentration. Namely, in the HC- (electrostatically) dominated regimes of large (small) solute sizes, salt increment reduces (raises) the splitting length $\sigma$ towards the limit value of $\sigma\sim d/2$.  Fig.~\ref{fig6}(c) also shows that below a characteristic packing fraction, the splitting length $\sigma$ vanishes, implying that the SCDH formalism converges to the CCDH approach. Most importantly, in Fig.~\ref{fig6}(d), one notes that the excess energy curves computed with the three different splitting potentials are considerably close. This indicates that the specific choice of the splitting scheme does not play a substantial role in the calculation of the bulk thermodynamic functions.

\subsection{Overscreening and underscreening}

We evaluate here the accuracy of our formalism in predicting the deviation of the macromolecular interaction range from the DH length~\cite{exp,safran,andelman}. Due to the high order of the moment integral $I^{(4)}_i$ in Eq.~(\ref{cor16}), in the calculation of the dressed charge, the uncertainties originating from the numerical inversion of the Fourier-transformed Green's function~(\ref{eq38}) induce substantially large errors. In order to avoid this complication, we evaluated the screening parameter~(\ref{cor20}) within the quadratic splitting scheme of Eq.~(\ref{eq53}) enabling the analytical inversion of Eq.~(\ref{eq38}) as
\be
\label{eq55}
G_{{\rm l},0}(r)=\frac{\ell_{\rm B}}{\gamma r}\left[e^{-\kappa_-r}-e^{-\kappa_+r}\right],
\ee
where $\gamma=\sqrt{1-4\kappa_0^2\sigma^2}$ and $\kappa_\pm=\sqrt{1\pm\gamma}/(\sqrt2\sigma)$.

\begin{figure}
\includegraphics[width=1.\linewidth]{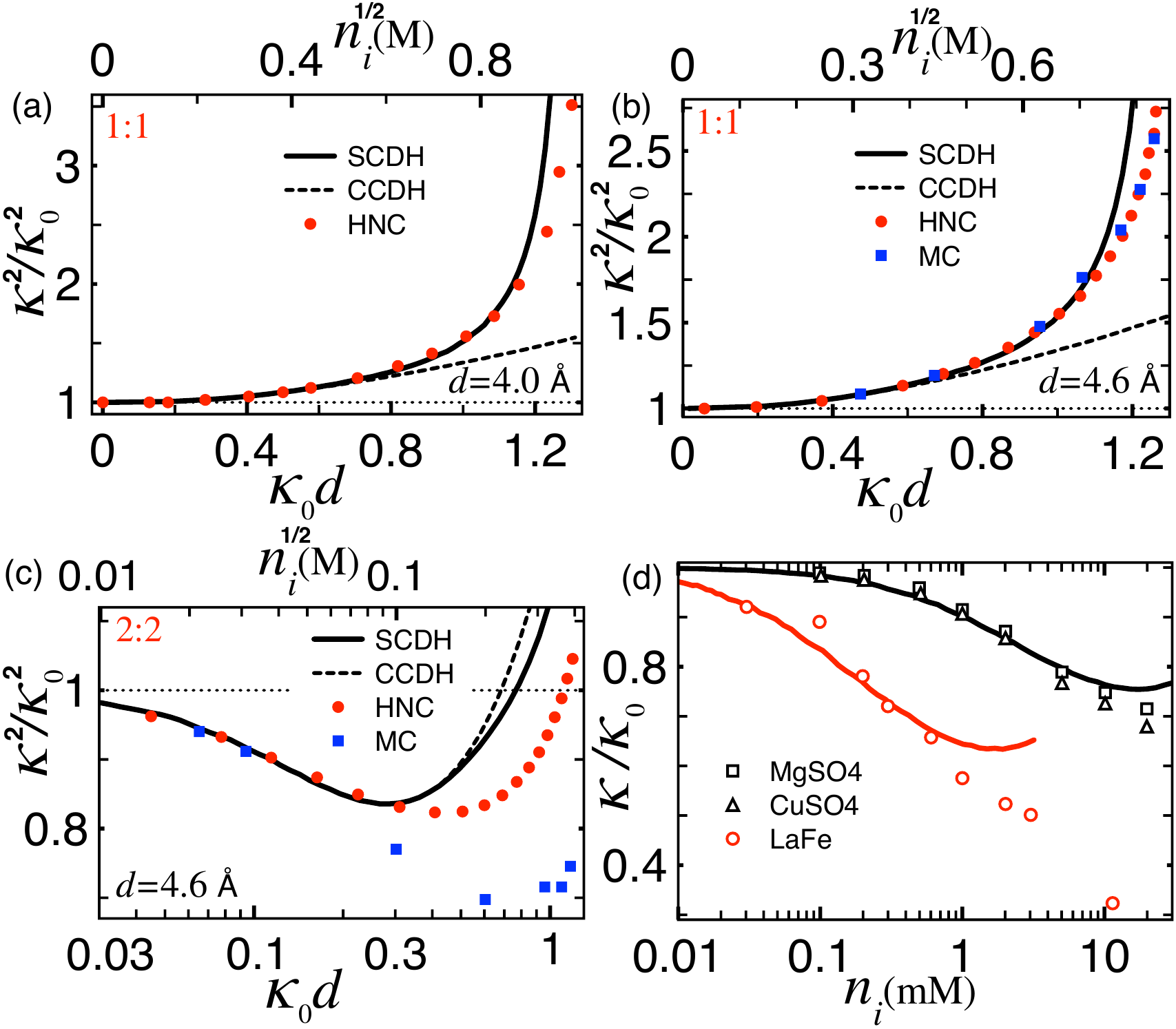}
\caption{(Color online) (a)-(c) Dimensionless screening parameter against concentration at the HC sizes and liquid compositions indicated in the legends. Solid curves: SCDH prediction in Eq.~(\ref{cor20}). Dashed curves: CCDH result. Disks: HNC data from (a) Fig.1 of Ref.~\cite{AttardRev} ($T=300$ K and $\e_{\rm w}=78.5$) and (b)-(c) Figs.10-11 of Ref.~\cite{Kj2020} ($T=298$ K and $\e_{\rm w}=78.4$). Blue squares: MC data of Ref.~\cite{Kj2020}. (d) Experimental data of Ref.~\cite{exp} (symbols) and SCDH prediction (curves) for the divalent ${\rm MgSO}_4$ and ${\rm CuSO}_4$ solutions ($d=3.2$ {\AA}), and the trivalent ${\rm LaFeCN}_6$ electrolyte ($d=6.3$ {\AA}) at the temperature $T=298$ K and dielectric permittivity $\e_{\rm w}=78.4$.}
\label{fig7}
\end{figure}

In Figs.~\ref{fig7}(a)-(b), we display the screening parameter of monovalent liquids at two different ion sizes. One sees that the CCDH formalism can reproduce the overscreening effect ($\kappa>\kappa_0$) only perturbatively. However, despite being based on the leading-order perturbative solution of Eq.~(\ref{cor19}), the SCDH prediction~(\ref{cor20}) can accurately account for the substantial increment of the dimensionless screening parameter by salt addition up to $n_i\sim0.5$ M for $d=4.6$ {\AA}, and up to $n_i\sim0.8$ M for $d=4.0$ {\AA}. Beyond these concentrations, the agreement of the SCDH prediction with the HNC and MC data is only qualitative.

The overscreening effect observed in Figs.~\ref{fig7}(a)-(b) is induced by the HC size of each ion keeping the surrounding counterion cloud away from the central charge~\cite{Kj1,Kj2,Kj1995,Kj2020}. The resulting void attenuates the shielding of the electric field created by the ion charge, enhancing in turn the overall screening ability of the electrolyte.

Fig.~\ref{fig7}(c) displays the reverse case of divalent solutions characterized by strong counterion attraction favoring the formation of ion pairs. The pair formation partially neutralizing the liquid lowers its charge strength and reduces its screening ability~\cite{Kj1,Kj2,Kj1995,Kj2020}. In agreement with HNC data, this underscreening effect ($\kappa<\kappa_0$) is reproduced by the CCDH and SCDH formalisms up to $n_i\sim10$ mM. However, the prediction of these three approaches deviate from the MC data at lower concentrations.  As a result, the reversal of the underscreening trend displayed by the MC data is significantly exaggerated by our formalisms as well as by the HNC approach.

Finally, in Fig.~\ref{fig7}(d), we confront the screening parameter~(\ref{cor20}) of the  SCDH formalism with the experimental data of Ref.~\cite{exp} for the divalent ${\rm MgSO}_4$ and ${\rm CuSO}_4$ solutions, and the trivalent ${\rm LaFeCN}_6$ electrolyte. The figure shows that by using the hydrated ion size as the only fitting parameter (see the caption), our approach can reproduce with reasonable accuracy the underscreening effect displayed by the experimental data for dilute salt. At larger concentrations, the departure from the experimental data accompanied with the exaggerated reversal of the curve trend is similar to that observed in Fig.~\ref{fig7}(c). The access to the corresponding strong salt regime requires the use of calculation schemes beyond the first-order cumulant closure presented in Sec.~\ref{gencum}. 

\SB{At this point, we note that the effect of salt-induced dielectric decrement~\cite{BuyukDielDec} may also contribute to the discrepancy between the theoretical curves and the experimental data in the large concentration regime of Fig.~\ref{fig7}(d). In future works, the corresponding feature may be taken into account by using salt-dependent dielectric permittivities~\cite{NetzMC2,Buyuk2020} or via the incorporation of explicit solvent into the present formalism~\cite{BuyukDielDec}.}

\section{Conclusions}

In this article, we presented an ion size-augmented self-consistent formalism of bulk electrolytes exploiting the asymmetric treatment of the distinct ionic interaction modes. Within this hybrid approach, the long-range interactions are taken into account via a first order cumulant expansion assuming moderate fluctuations of the long-wavelength modes. Then, the short-range interactions are incorporated via a virial treatment that allows to avoid the WC-level gaussian approximation. 

These distinct ranges are separated by the splitting parameter $\sigma$ satisfying the newly derived variational identity~(\ref{eq13}). This identity also valid for inhomogeneous ion solutions can be readily used in future works to generalize the present formalism to the case of confined electrolytes. Moreover, the corresponding variational scheme can be directly extended to more general splitting potentials including multiple splitting parameters $\sigma_i$.

For monovalent electrolytes with typical hydrated ion sizes $d\sim4$ {\AA}, the asymmetric treatment of the short- and long-range ion interactions allows the accurate prediction of \SB{the HC-dominated osmotic pressures, the electrostatic correlation-driven internal energies, and the ionic pair distributions governed by the competition between electrostatic and HC interactions} up to the molar concentration regime. This is a substantial upgrade of our earlier CCDH formalism. In the case of multivalent ions, the quantitative accuracy of the SCDH formalism is limited to submolar concentrations. However, even in the vicinity of the molar concentration range, the SCDH approach improves significantly the qualitative agreement of the CCDH formalism with MC simulations.

Within the framework of our functional integral theory, we reproduced as well the characteristic Eq.~(\ref{cor20}) of the screening parameter previously derived by the dressed-ion formalism~\cite{Kj1995}. The leading-order perturbative solution of this equation has been compared with HNC and MC results as well as with experimental data. We found that for monovalent ions with typical sizes, the SCDH approach can accurately reproduce the overscreening effect close to the molar concentration regime. In the case of multivalent electrolytes, the accuracy of our formalism in predicting the underlying underscreening effect is limited to strictly dilute salt concentrations. With the aim to extend our results to strong salt conditions, we are currently developing upgraded versions of our formalism \SB{based on high order closures of the SD Eq.~(\ref{eq21}) and incorporating explicit solvent~\cite{BuyukDielDec}}. These improvements will be reported in future works.

\smallskip
\textbf{Data availability}

The Author declares that the parameters and data supporting the graphically illustrated results are available within the article. The bibliographic references of the HNC, MC, and experimental data taken from the literature are cited in the caption of each figure.

\smallskip
\textbf{Conflict of interest}

The author declares no competing interests.


\begin{thebibliography}{99}
\bibitem {rem1} In Ref.~\cite{NetzMC}, the MC simulation data has been provided in terms of the dimensionless ion size and packing fractions (the lower axis in Figs.~\ref{fig5} and~\ref{fig6}). In the plots of Figs.~\ref{fig5} and~\ref{fig6}, these parameters have been converted to the physical ion sizes (see the legends) and salt concentrations (the upper axis) for monovalent ions ($q_i=1$) and the model parameters indicated in the caption.
\end{thebibliography}
\end{document}